\documentclass[]{mn2e}
\usepackage{psfig, epsf, epsfig}
\title[Dark galaxies or tidal debris?]{Dark galaxies or tidal debris? 
       Kinematical clues to the origin of massive isolated HI clouds}
\author[K. Bekki,  V. A. Kilborn, and B. S. Koribalski]
       {Kenji Bekki${}^1$\thanks{E-mail: bekki@bat.phys.unsw.edu.au}, 
        B\"arbel S. Koribalski${}^2$, and
        Virginia A. Kilborn${}^{2,3}$ \\
       ${}^1$School of Physics, University of New South Wales,
              Sydney 2052, NSW, Australia\\
       ${}^2$Australia Telescope National Facility, CSIRO, P.O. Box 76,
              Epping NSW 1710, Australia\\
       ${}^2$Centre for Astrophysics and Supercomputing,
              Swinburne University of Technology, Mail H39, PO Box 218,
              Hawthorn, VIC 3122, Australia \\}
  
\begin{document}

\date{Accepted, Received 2005 May 13; in original form }

\pagerange{\pageref{firstpage}--\pageref{lastpage}} \pubyear{2005}

\maketitle

\label{firstpage}

\begin{abstract}
An extended HI cloud (VIRGOHI\,21) with an HI mass of $\sim10^8$ M$_{\odot}$ 
and no apparent optical counterpart was recently discovered 
in the Virgo cluster. 
In order to understand the origin of
physical properties of apparently isolated HI clouds
like the VIRGOHI21, we numerically investigate
physical properties of  tidal HI debris that were formed
by galaxy-galaxy interaction in clusters of galaxies. 
Our hydrodynamical simulations 
demonstrate that tidal debris with total HI masses of $10^8-10^9$ M$_{\odot}$ 
can have (1) wide HI velocity widths ($>200$ km\,s$^{-1}$), (2) a small mass 
fraction of stars ($\sim$10\%), and (3) a mean $B$-band surface brightness of 
the stellar components fainter than 30 mag\,arcsec$^{-2}$. 
These results suggest that the VIRGOHI21,
which lies at a projected distance of $\sim$ 150 kpc from the one-armed,
HI-rich spiral galaxy M99 (NGC 4254), is tidal debris.
We propose that the comparison between the simulated 
and the observed velocity fields 
of HI clouds allows us to better understand their nature and origin
(e.g., whether they are just tidal debris or ``dark galaxies'' that have HI
gas only and are embedded by
dark matter halos). 
We also discuss the timescales for isolated HI gas 
to be evaporated by thermal conduction of the hot intracluster gas.

\end{abstract}


\section{Introduction}

Detections of HI structures without apparent stellar counterparts are not
uncommon in the outskirts of galaxies and within galaxy groups.
The Galactic high velocity clouds (HVCs) and the prominent HI tidal streams 
tracing the interaction between the Milky Way and the Magellanic Clouds are 
the closest examples. Another prime example is the nearby, HI-rich group 
consisting of the galaxies M\,81, M\,82, and NGC~3077 which are connected by 
a beautiful network of HI filaments and streams (Yun et al. 1994). In most 
cases, the location, structure and velocity field of the apparently star-less
HI clouds, with respect to the nearby galaxies, clearly indicates their tidal 
or collisional origin. Hibbard et al. (2001) show a large number of examples 
collected as part of the `Rogues Gallery of Galaxies': there are one-sided 
tidal tails stretching out to $\sim$100 kpc (e.g., Appelton et al. 1987, Rots 
et al.  1990, Clemens et al. 1999), HI rings with 100--200 kpc diameter (e.g.,
Schneider et al. 1989, Malphrus et al. 1997), etc. On the other hand, isolated 
HI clouds appear to be extremely rare as shown by the HI Parkes All-Sky Survey 
(Koribalski et al. 2004). The most isolated HI cloud discovered, HIPASS 
J0731--69 (Ryder et al. 2001), lies at a projected distance of 180 kpc from 
the asymmetric spiral galaxy NGC~2442 within the NGC~2434 galaxy group.  
Bekki et al. (2005) suggested a possible formation scenario 
that this massive cloud is  likely to be 
the high column density peak of a much larger HI structure.

Davies et al. (2004) recently discovered several HI clouds in the Virgo 
cluster. At least one of these clouds, VIRGOHI\,21 ($v_{\rm hel} \sim 2000$ 
km\,s$^{-1}$), has no apparent optical counterpart and is, on the basis of
its relatively wide HI spectrum, interpreted as a dark matter halo by Minchin 
et al. (2005). While the projected distance of VIRGOHI\,21 to the one-armed 
spiral galaxy M\,99 (NGC~4254, $v_{\rm sys} \approx 2400$ km\,s$^{-1}$) may 
appear large (25\arcmin\ or $\sim$150 kpc at a distance of 20 Mpc), it is by 
no means exceptional as indicated above. Detailed VLA HI observations of M\,99, 
the brightest spiral in the Virgo cluster, by Phookun et al. (1993) reveal a 
large amount of gas ($\sim2.3 \times 10^8$ M$_{\odot}$) at peculiar velocities,
stretching at least 11\arcmin\ toward the northwest (in the direction of 
VIRGOHI\,21) in form of a faint clumpy trail. The total HI mass measured in 
the M\,99 system by Phookun et al. (1993) is $7.6 \times 10^9$ M$_{\odot}$, 
typical for its morphological type (Sc) and luminosity; the HI mass to light 
ratio is $\sim$0.1. Using HIPASS we measure an HI flux density of $80 \pm 2$ 
Jy\,km\,s$^{-1}$ for M\,99 (HIPASS J1218+14), in agreement with the value 
obtained by Phookun et al. (1993), while the VIRGOHI\,21 cloud is too faint 
for detection. 
M\,99 has the possible dynamical mass
of $\sim1.6 \times 10^{11}$ M$_{\odot}$
and is located in the outskirts of the Virgo cluster, about 
3.7 degrees ($\sim$1.3 Mpc projected distance) northwest of the giant 
elliptical M\,87 and just outside the region of strongest X-ray emission.
Minchin et al. (2005) discount tidal origins for the cloud,
though the position of the cloud on the outskirts of a large cluster, and the
proximity of the HI disturbed spiral NGC 4254 (described above)
warrant a closer investigation of the origins of the cloud.

The purpose of this paper is thus to investigate whether the tidal
debris scenario is a viable explanation for the apparently
isolated massive HI clouds
like VIRGOHI21.  Based on the numerical simulations on the
formation of tidal debris in interacting galaxies in a  
cluster of galaxies, 
we particularly discuss (1) whether the observed physical
properties of apparently isolated HI clouds, 
such as wider velocity width ($\Delta V
\sim 200$ km s$^{-1}$) of the clouds and no visible counterparts, can
be reproduced by the tidal debris scenario and (2) what are key
observations that can discriminate between the two scenarios
(i.e., ``dark galaxy'' vs ``tidal debris'').
We thus focus on the tidal stripping scenario, though
ram pressure stripping can be also associated with the origin
of isolated HI clouds (e.g., Vollmer et al. 2005).
The present numerical results are discussed in a more general way
without comparing the results with observations of specific targets
(e.g., VIRGOHI21), though the numerical model is more reasonable for
galaxy evolution in the Virgo cluster.  
The present results therefore  can be useful in interpreting the
observational results of 
other (apparently) isolated HI objects such as HIPASS J0731-69
(Ryder et al. 2001).

\begin{table*}
\centering
\begin{minipage}{185mm}
\caption{Model parameters and results}
\begin{tabular}{cccccccccc}
(1)&(2)&(3)&(4)&(5)&(6)&(7)&(8)&(9)&(10)\\
Model no.                                  &
$M_{\rm d}$ ($\times 10^{10}$ M$_{\odot}$) &  
$r_{\rm p}$  (kpc)                         & 
${\theta}_{\rm d}$ (degrees)               &  
$m_{2}$                                    & 
$R_{\rm ini}$  (kpc)                       &  
$M_{\rm HI}$ ($\times 10^9$ M$_{\odot}$)   & 
$f_{\rm s}$                                &
${\mu}_{\rm B}$ (mag/arcsec$^{2}$)         &
$r_{\rm sep}$ (kpc)                        \\ 
M1 & 0.6 &  22 &  45 & 3   & 350 & 0.17 & 0.14 & 33.4 & 532 \\
M2 & 0.6 &  6  &  45 & 3   & 700 & 0.07 & 0.25 & 33.7 & 171 \\
M3 & 0.6 &  22 &  45 & 3   & 700 & 0.43 & 0.57 & 30.2 & 259 \\
M4 & 6.0 &  18 & 150 & 0.3 & 700 & 0.84 & 0.30 & 30.6 & 310 \\
\end{tabular}
\end{minipage}
{\bf Notes:} Cols. (1--6) model parameters, Cols. (7--9) total gas mass, mass 
 fraction of stars, and average $B$-band surface brightness, respectively, 
 within a 100 kpc box as shown in Fig.~1, Col. (10) separation of the two 
 interacting galaxies after 2 Gyrs.
\end{table*}

\begin{figure}
\psfig{file=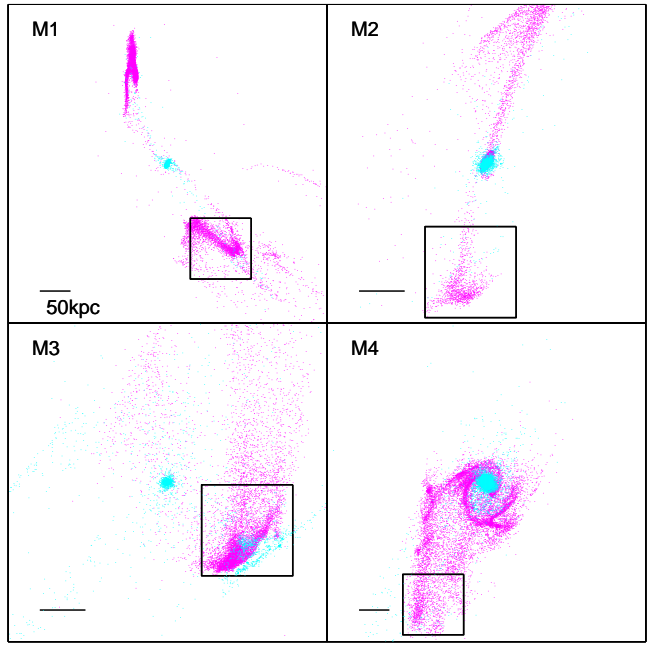,width=8.5cm}
\caption{
 Distribution of the stars (cyan) and gas (magenta) in an interacting galaxy 
 pair within the Virgo cluster after $\sim$2 Gyr orbital evolution for our 
 models M1 to M4.
 The distributions are projected onto the $x-y$ plane with the frame center 
 coincident with the center of the disk galaxy. The bar in the lower left 
 corner of each panel represents a scale of 50 kpc. The inserted box has a 
 size of 100 kpc and indicates the location of relatively isolated tidal 
 debris for which structural and kinematical properties are described in 
 the text and Table~1. 
The companion, which was modelled as a point mass, is outside
the box; the separation between the two interacting galaxies
after 2 Gyrs is given in Table 1, (Col. 10).
Note that M1 and M2 models show leading and trailing HI tails, which
have different shapes and densities.}
\label{Figure. 1}
\end{figure}

\section{The Model}

Details on the disk galaxy models and the external gravitational
potential of clusters of galaxies adopted in the present study have
already been described in Bekki et al. (2005) and Bekki et al. (2003),
respectively, so we give only a brief review here. We investigate the
dynamical evolution of stellar and gaseous components in an
interacting pair of late-type disk galaxies orbiting the center of the
Virgo cluster by using TREESPH simulations (Bekki et al. 2002). A
late-type disk galaxy with a total mass of $M_{\rm t}$, a total disk mass
of $M_{\rm d}$, and a stellar disk size of $R_{\rm s}$ is assumed to be
embedded in a massive dark matter halo with a universal `NFW' profile
(Navarro et al. 1996) with a mass of $M_{\rm dm}$, and an exponential
stellar distribution with a scale length ($R_0$) of $0.2 R_{\rm s}$. 
The mass ratio of the dark matter halo to the stellar disk is set to be 
9 for all models (i.e., $M_{\rm dm}/M_{\rm t} = 0.9$). The galaxy is 
assumed to have an extended HI gas disk with an initial size ($R_{\rm g}$) 
of $2 \times R_{\rm s}$, which is consistent with observations (e.g., 
Broeils \& van Woerden 1994). An isothermal equation of state is used 
for the gas with a sound speed of 5.8 km\,s$^{-1}$ (corresponding to 
2500 K). The adopted $R_0-M_{\rm d}$ relation for disk models with 
different masses is described as:
\begin{equation}
R_0=3.5 {(\frac{M_{d}}{6 \times 10^{10} {\rm M}_{\odot}})}^{0.5} \, {\rm kpc},
\end{equation}
which is consistent both with the observed scaling relation for bright disk 
galaxies (Freeman 1970) and with the Galactic structural parameters (Binney 
\& Tremaine 1980).

The initial position (${\bf x}_{i}$, $i=1,2$) and velocity (${\bf v}_{i}$) of 
two interacting galaxies with respect to the cluster center is described as:
\begin{equation}
  {\bf x}_{i} = {\bf X}_{\rm g} + {\bf X}_{i} 
\end{equation}
and
\begin{equation}
  {\bf v}_{i} = {\bf V}_{\rm g} + {\bf V}_{i},
\end{equation}
respectively, where ${\bf X}_{\rm g}$ (${\bf V}_{\rm g}$) are the position 
(velocity) of the center of mass of an interacting pair with respect to the 
cluster center and ${\bf X}_{i}$ (${\bf V}_{i}$) is the location (velocity)  
of each galaxy in the pair with respect to their center of mass. 
${\bf X}_{i}$ and ${\bf V}_{i}$ are determined by the orbital eccentricity
($e$), the pericenter distance ($r_{\rm p}$), and the mass ratio of the two 
($m_2$). Only one of the two interacting galaxies is modeled as a fully 
self-consistent disk described above whereas the other is modeled as a 
point mass with the total mass of $m_{2} \times M_{\rm t}$.
Accordingly, the total mass of a pair in a model is $(1+m_2) \times M_{t}$,
which corresponds to $2.4 \times 10^{11} {\rm M}_{\odot}$ for the
model M1 (See Table 1).
The self-consistent disk model is inclined by ${\theta}_{\rm d}$ 
(degrees) with respect to the orbital plane of the interacting pair.

To give our model a realistic radial density profile for the dark matter 
halo of the Virgo cluster and thereby determine ${\bf X}_{\rm g}$ and 
${\bf V}_{\rm g}$, we base our model on both observational studies of the 
mass of the Virgo cluster (e.g., Tully \& Shaya 1984) and the predictions 
from the standard cold dark matter cosmology (NFW). The NFW profile is 
described as:
\begin{equation}
  {\rho}(r)=\frac{\rho_{0}}{(r/r_{\rm s})(1+r/r_{\rm s})^2},
\end{equation}
where $r$, $\rho_{0}$, and $r_{\rm s}$ are the distance from the center
of the cluster, the central density, and the scale-length of the dark halo,
respectively. The adopted NFW model has a total mass of $5.0 \times 10^{14}$
M$_{\odot}$ (within the virial radius) and $r_{\rm s}$ of 161\,kpc. 
The center of the cluster is always set to be ($x$,$y$,$z$) = (0,0,0) whereas  
${\bf X}_{\rm g}$ is set to be ($x$,$y$,$z$) = ($R_{\rm ini}$, 0, 0). 
${\bf V}_{\rm g}$ is set to be ($v_{\rm x}$,$v_{\rm y}$,$v_{\rm z}$) = 
(0, $f_{\rm v} V_{\rm c}$, 0), where $f_{\rm v}$ and $V_{\rm c}$ are the 
parameters controlling the orbital eccentricity (i.e, the larger $f_{\rm v}$ 
is, the more circular the orbit becomes) and the circular velocity of the 
cluster at $R$ = $R_{\rm ini}$, respectively. The orbital plane of the pair  
is assumed to be inclined by 30 degrees with respect to the $x-y$ plane for 
all models. Thus $R_{\rm ini}$ and $f_{\rm v} $ are the two key parameters 
for the orbital evolution of the pair. We show the results of the models with
$f_{\rm v}=0.5$ and $e=1.5$ (high-speed, hyperbolic encounters appropriate 
for cluster galaxy interaction) in the present study.

Although we investigated a large number of models, we only show the results 
of four representative models which produce {\it relatively isolated gas 
clouds (i.e., tidal debris) without many stars}. The model parameters and 
the resulting physical properties 
of the tidal debris are summarized in Table~1: $M_{\rm g}$ (column 7), 
$f_{\rm s}$ (8), ${\mu}_{\rm B}$ (9), and $r_{\rm sep}$ (10) describe the 
total gas mass within a 100 kpc box within which relatively isolated tidal 
debris can be seen for each model (see Fig.~1), the mass fraction of stars 
within the debris, the $B$-band surface brightness of stars averaged over the 
100 kpc box, and the distance of two interacting galaxies in each panel of 
Fig.~1, respectively. 
The occurrence of single or double tails depends strongly on
the orbits and mass ratios.
We emphasize the results of M4, because this model 
suggests that a past interaction between NGC~4254 and a galaxy (that can not 
be specified in the present study) can be responsible both for the origin of
VIRGOHI\,21 and for the morphological properties of NGC~4254 (e.g., the
strong one-armed spiral structure and the one-sided tidal plume of gas).
 
The $B$-band mass-to-light ratio ($M_{\rm d}/L_{\rm B}$, where $L_{\rm B}$ 
is the $B$-band luminosity of a disk) is assumed to be four for estimating 
the surface brightness of tidal tails and debris. We mainly investigate 
column density distributions and velocity fields of tidal debris for 
different viewing angles (${\theta}_{i}$), i.e. the angle between the 
$z$-axis and the line-of-sight that is always within the $y-z$ plane.

\begin{figure}
\psfig{file=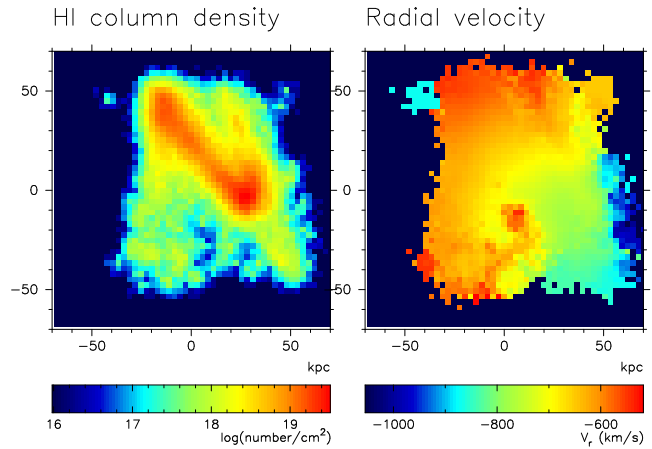,width=8.5cm}
\caption{ 
 Gas column density distribution (left) and velocity field (right) for model 
 M1. The cell size is 2.8 kpc and the viewing angle (${\theta}_{\rm i}$) is 
 $45^{\circ}$. 
}
\label{Figure. 2} 
\end{figure}

\begin{figure}
\psfig{file=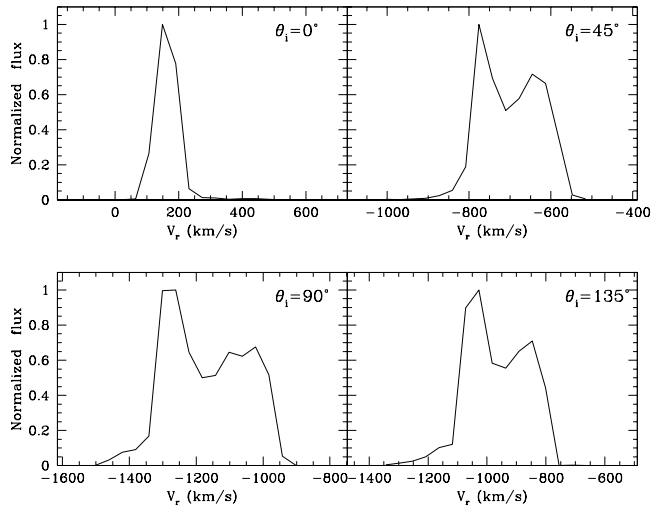,width=8.5cm}
\caption{ 
 Simulated spectra of the gas in model M1. The normalized flux at each 
 velocity $V_{\rm r}$ is derived by estimating the total mass of gas 
 particles with radial velocities $\sim V_{\rm r}$.
}
\label{Figure. 3}
\end{figure}

\begin{figure}
\psfig{file=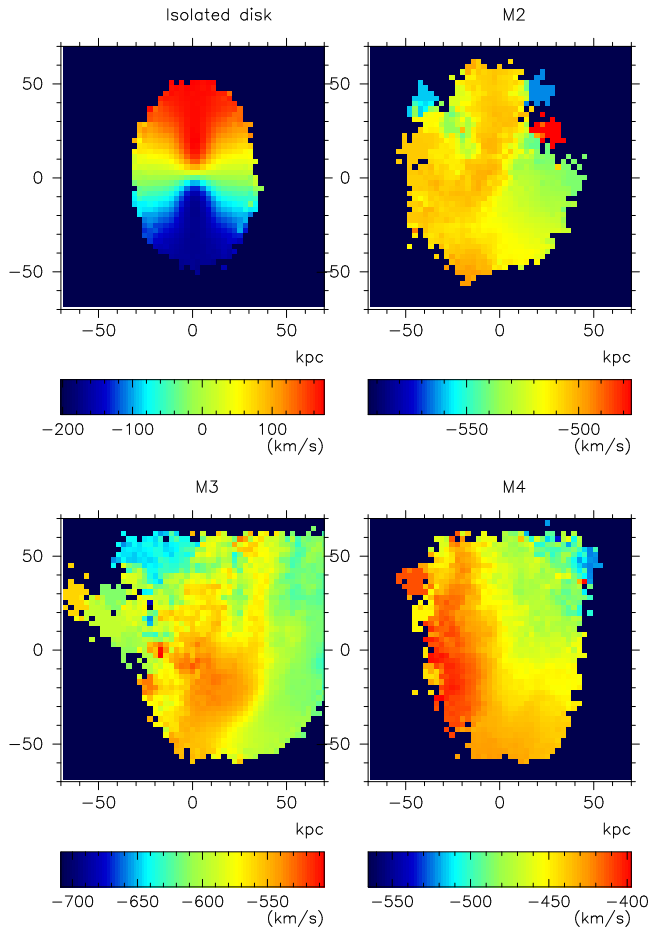,width=8.5cm}
\caption{ 
 Simulated velocity fields for the gas in the initial disk (upper left; 
 inclined for comparison), and the tidal debris resulting from our models:
 M2 (upper right), M3 (lower left), and M4 (lower right). The color bar
 at the bottom of each panel indicates the range of velocities ($\sim$100
 to 200 km\,s$^{-1}$) in the simulated debris.
}
\label{Figure. 4}
\end{figure}

\section{Results}

Fig.~1 summarises the projected mass distributions of stars and gas after 
$\sim$2 Gyr dynamical evolution of interacting galaxies for the four models 
(M1, M2, M3, and M4).  Stars and gas tidally stripped by the galaxy-galaxy 
interaction (and by the cluster tidal field) can not return back to the host 
galaxies owing to the stronger cluster tidal field (e.g., Mihos 2004). As a 
result of this, they form either tidal debris an the tip of the tidal tails 
or faint `tidal bridges' connecting the host galaxies. The tidal debris can 
have large HI masses ($10^8 - 10^9$ M$_{\odot}$) and be well detached from 
their host galaxies. The interaction partners have separated 
($\sim 100$ kpc) 
from the tidal tails and debris which are then observed as relatively 
isolated HI clouds.

Tidal stripping is much more efficient in gas than in stars, because,
in our model,
the initial gas distributions in the disks are two times more extended
than the stellar ones.  Therefore, the mass fraction of stars can be
small ranging from 14\% (M1) to 57\% (M3) within the `isolated HI
clouds' (shown within the box of each model in Fig.~1).  Stripped
stars can be very diffusely distributed and consequently the mean
$B$-band surface brightness (${\mu}_{\rm B}$) within the 100 kpc box
is 33.7 mag\,arcsec$^{-2}$ (in model M1). The morphological properties of
the gas and stars, the stellar mass fraction, and ${\mu}_{\rm B}$ of
the stars in the tidal debris (or `isolated HI clouds') are quite
diverse and depend on projections. For example, ${\mu}_{\rm B}$ is
33.7 mag\,arcsec$^{-2}$ for model M2 whereas it is 30.2 mag\,arcsec$^{-2}$ 
for model M3. The derived very faint surface brightness suggests that it 
is almost impossible to detect any optical counterparts of the tidal
debris with current large telescopes and reasonable exposure times.

Fig.~2 shows the column density distribution and velocity field of the 
tidal debris in model M1. It is clear from this figure that (1) the gaseous 
distribution is quite irregular and inhomogeneous with local column densities 
ranging from $\sim 10^{17}$ to $\sim 10^{19}$ cm$^{-2}$, (2) there is a 
strong velocity gradient between the upper left and the lower right parts 
of the debris, but the overall velocity field is appreciably irregular, 
and (3) there is no clear sign of global rotation. Result (1) implies that
the observed morphology of the tidal debris depends strongly on the column
density limit of the observations.

Fig.~3 shows that the simulated spectra (i.e., the integrated $V_{\rm r}$
distribution) of the gaseous debris can have a wide velocity range ($>$200 
km\,s$^{-1}$) for some viewing angles. But unless the gas is shown to be
rotating, the velocity width of the HI spectrum alone does not lead to an
estimate of the cloud mass. Streaming motions within the tidal debris are
responsible for the wide velocity width of the gas in the present models.  
This kind of wide velocity width in HI gas has been  already observed 
for the relatively isolated massive HI cloud 
discovered by HIPASS within the NGC~2434 galaxy group (Ryder et al. 2001).

Fig.~4 illustrates how much the simulated velocity fields of the gas in our
models differ from that of a regular rotating gas disk. It is a generic result 
of the present study that the velocity fields of tidal debris do not resemble 
the typical `spider diagrams' which are characteristic of regular HI kinematics
in disk galaxies that are supported by rotation against gravitational fields 
made by baryonic and dark matter of the galaxies. Velocity fields are the key 
observational tools to help us determine whether gas clouds are (unbound) 
tidal debris or a self-gravitating systems embedded within a massive dark 
matter halo.

\begin{figure}
\psfig{file=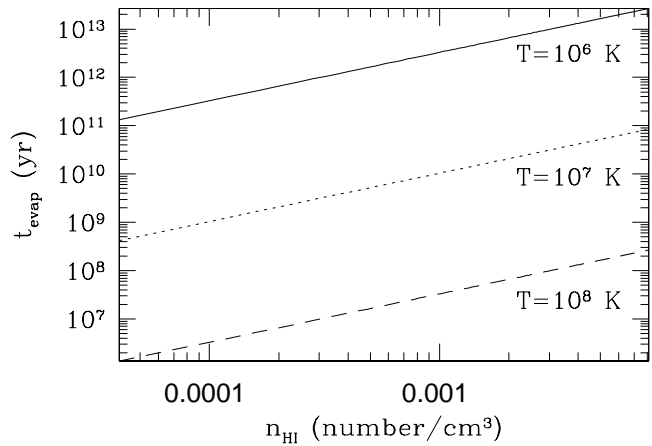,width=8.5cm}
\caption{ 
The dependences of the evaporation time scale of HI gas ($t_{\rm evap}$)
on HI densities ($n_{\rm HI}$) for different temperature ($T$) of hot gas in
a cluster of galaxies:
$T=10^6$ (solid), $T=10^7$ (dotted), and $T=10^8$ K (dashed).
}
\label{Figure. 5}
\end{figure}

\section{Discussions and Conclusions}

We have shown that tidal debris formed from interacting galaxies in clusters 
(1) can have a wide velocity width ($>$200 km\,s$^{-1}$), (2) show 
${\mu}_{\rm B}$ of stars fainter than 30 mag\,arcsec$^{-2}$, and (3) are 
located far from ($>$100 kpc) their progenitors. Our results suggest that 
the HI cloud VIRGOHI\,21 which lies at a distance of $\sim$150 kpc from 
the one-armed spiral M\,99 (NGC~4254) is likely to be tidal debris rather 
than a `dark galaxy'. We stress that a detailed kinematical study of the 
HI gas in the extended region around M\,99, and of the region between 
VIRGOHI\,21 and M\,99 in particular, is necessary to fully understand the 
origin of the peculiar HI gas.

If VIRGOHI\,21 is indeed tidal debris, how long can it be observed as cold 
HI gas in the Virgo cluster where it can be easily evaporated by thermal 
conduction of the hot intracluster gas? Cowie \& McKee (1977) investigated 
the time scale ($t_{\rm evap}$) within which an isolated spherical gas cloud 
embedded by hot tenuous gas can be evaporated and estimated $t_{\rm evap}$ 
as:
\begin{equation}
  t_{\rm evap}=3.3 \times 10^{20} n_{\rm cl} {R_{\rm cl}}^2 T^{-2.5} \,
  {\rm yr},
\end{equation}
where $n_{\rm cl}$, $R_{\rm cl}$, and $T$ are the mean density of a gas cloud, 
the size of the gas distribution in pc, and the temperature of the surrounding 
hot gas, respectively. Here the value of the Coulomb logarithm of $\ln \Lambda$
is assumed to be 30 (Cowie \& McKee 1977). Using the present results for model 
M1, we can estimate $t_{\rm evap}$ as:
\begin{eqnarray}
  t_{\rm evap}=3.7 \times 10^7 (\frac{n_{\rm cl}}{3.2 \times 10^{-4} 
  {\rm cm}^{-3}}) \times
  {(\frac{R_{\rm cl}}{10^5 \rm pc})}^{2} \nonumber \\
  \times {(\frac{T}{3.5 \times 10^5 {\rm K}})}^{-2.5} \, {\rm yr},
\end{eqnarray}
where we adopt the reasonable value of 3 keV for the temperature of the hot 
gas in the central region of the Virgo cluster (e.g., Matsumoto et al. 1996)
and use the mean gas density within the 100 kpc box in model M1. $n_{\rm cl}$ 
varies within the tidal debris, and $T$ around 
gas clouds can also vary depending on the location of the clouds within a 
cluster owing to the radial $T$ gradient (e.g., Matsumoto et al. 1996). We 
therefore estimate $t_{\rm evap}$ for different $n_{\rm cl}$ and $T$; the 
results are shown in Fig.~5.  It is clear that low density HI clouds with
$n_{\rm cl}$ = $10^{-4}$ to $10^{-3}$ cm$^{-3}$ can be evaporated by thermal 
conduction well within $10^{10}$ yr in cluster environments with $T$ =
$10^7 - 10^8$ K: these clouds would be difficult to observe in clusters a 
few Gyr after their formation.

Recently discovered massive HI clouds with no apparent optical counterparts 
have provided new clues to several problems of extragalactic astronomy such 
as the origin of HVCs and intergalactic star-forming regions (e.g., Kilborn 
et al. 2000, Ryder et al. 2001, Ryan-Weber et al. 2004, Kilborn et al. 2005).
Although previous numerical studies have tried to reproduce the observed 
structural properties of relatively isolated HI clouds (e.g., Bekki et al. 
2005), they did not discuss the kinematical properties of the clouds 
extensively. Simulated velocity fields as well as particle/mass distributions 
are necessary for comparison with the observed gas kinematics in galaxies 
and surrounding material to better understand galaxy interactions in various
environments.

\section*{Acknowledgments}
KB  acknowledges the financial support of the Australian Research 
Council throughout the course of this work.
The numerical simulations reported here were carried out  
at Australian Partnership for Advanced Computing (APAC).

\end{document}